\documentclass[superscriptaddress,dvipsnames,nofootinbib,amsmath,amssymb,prd,notitlepage,prd,showpacs,twocolumn]{revtex4-1}
\usepackage{color}
\usepackage{bm}
\usepackage{graphicx}
\usepackage{amsmath}
\usepackage{amssymb}
\usepackage{enumitem}
\usepackage{hyperref}
\usepackage{subfigure}
\usepackage{array}
\usepackage[english]{babel}
\usepackage{tensor}
\usepackage{comment}
\usepackage[normalem]{ulem}
\usepackage[dvipsnames]{xcolor}

\usepackage{fontawesome}

\def\be{\begin{equation}}
\def\ee{\end{equation}}
\def\ba{\begin{eqnarray}}
\def\ea{\end{eqnarray}}
\def\l{\left}
\def\r{\right}
\def\f{\frac}

\allowdisplaybreaks

\def\nn{\nonumber}

\def\ak{\alpha_K}
\def\ab{\alpha_B}

\newcommand{\Omegam}{\ensuremath{\Omega_{\mathrm{m},0}}}

\newcommand{\de}{\mathrm{d}}

\begin{document}

\title{Non-linear  power spectrum and forecasts for Generalized Cubic Covariant Galileon}

\author{Lu\'is Atayde}
\email{luisbbatayde@gmail.com}
\affiliation{
 Instituto de Astrofis\'ica e Ci\^{e}ncias do Espa\c{c}o, Faculdade de Ci\^{e}ncias da Universidade de Lisboa, Edificio C8, Campo Grande, P-1749016, Lisboa, Portugal }
 \author{Noemi Frusciante}
 \email{noemi.frusciante@unina.it}
 \affiliation{Dipartimento di Fisica ``E. Pancini", Universit\`a degli Studi  di Napoli  ``Federico II", Compl. Univ. di Monte S. Angelo, Edificio G, Via Cinthia, I-80126, Napoli, Italy}
\author{Benjamin Bose}
\email{ben.bose@ed.ac.uk}
\affiliation{Institute for Astronomy, University of Edinburgh, Royal Observatory, Blackford Hill, Edinburgh, EH9 3HJ, UK}
\affiliation{Basic Research Community for Physics e.V., Mariannenstraße 89, Leipzig, Germany}
 \author{Santiago Casas}
\email{casas@physik.rwth-aachen.de}
\affiliation{
Institute for Theoretical Particle Physics and Cosmology (TTK),
RWTH Aachen University, 52056 Aachen, Germany}
\author{Baojiu Li}
\email{baojiu.li@durham.ac.uk}
\affiliation{
 Institute for Computational Cosmology, Department of Physics, Durham University, South Road, Durham, DH1 3LE, UK}
 
\begin{abstract}
To fully exploit the data from next generation surveys, we need an
accurate modelling of the matter power spectrum up to non-linear scales. Therefore in this work we present the halo model reaction framework for the  Generalized Cubic Covariant Galileon (GCCG) model, a modified gravity model within the Horndeski class of theories which extends the cubic covariant Galileon (G3) by including power laws of the  derivatives of the scalar field in the K-essence and cubic terms.  We modify the publicly available software \texttt{ReACT} for the GCCG in order to obtain an accurate prediction of the non-linear power spectrum.  In the limit of the G3 model we compare the modified  \texttt{ReACT} code to $N$-body simulations and we find agreement within 5\% for a wide range of scales and redshifts.  We then study the relevant effects of the modifications introduced by the GCCG on the non-linear matter power spectrum.
Finally, we provide forecasts from spectroscopic and photometric primary probes by next generation surveys using a Fisher matrix method. We show that future data will be able to constrain at 1$\sigma$ the two additional parameters of the model at the percent level and that  considering non-linear corrections to the matter power spectrum beyond the linear regime is crucial to obtain this result.  
\end{abstract}

\date{\today}
\maketitle
\tableofcontents

\section{Introduction}

The observed late-time acceleration of the Universe remains one of the greatest mysteries in cosmology. The standard cosmological model ($\Lambda$CDM) assumes the theory of General Relativity (GR) to describe the space-time, with  the cosmological constant, $\Lambda$, being the source of the cosmic acceleration. This model has well known theoretical \cite{Weinberg:1988cp,Carroll:2000fy,Velten:2014nra,Joyce:2014kja} and observational \cite{Riess:2019cxk,Wong:2019kwg,Freedman:2019jwv,DiValentino:2020zio,Kuijken:2015vca,deJong:2015wca,Hildebrandt:2016iqg,DiValentino:2020vvd} shortcomings which motivated the search for models beyond GR, known as Modified Gravity (MG) models \cite{CANTATA:2021ktz}. 

Scalar-tensor theories are among these MG models and are characterized by an additional scalar field which is coupled to gravity through non-minimal and derivative couplings.  Many of these models belong to the  Horndeski and/or Galileon theories \cite{Horndeski:1974wa,Nicolis:2008in,Deffayet:2009wt,Deffayet:2009mn,Kobayashi:2011nu}. They have gained a lot of attention given the possibility to construct viable models to account for the late-time acceleration without a cosmological constant as well as to realize phantom dark energy equation-of-state free of ghosts. Additionally in the sub-class of Galileon models \cite{Creminelli:2017sry,Baker:2017hug,Ezquiaga:2017ekz}  satisfying the constraints on the speed of gravitational waves \cite{LIGOScientific:2017zic}, i.e. $c_{\rm GW}=1$, there are models for which it has been shown to have a better fit to cosmological data than  $\Lambda$CDM \cite{Peirone:2019aua,Frusciante:2019puu}. An example is the Generalized Cubic Covariant Galileon model \cite{DeFelice:2011bh} (GCCG), for which a clear preference over $\Lambda$CDM is found when the Planck cosmic microwave background (CMB) temperature and polarization  data are used in the analysis \cite{Frusciante:2019puu}. Moreover  for this model the value of the today’s Hubble parameter, $H_0$,  is found to be consistent with its determination from Cepheids at 1$\sigma$ \cite{Frusciante:2019puu}, resolving the famous tension of the cosmological standard model \cite{DiValentino:2020zio}.  In details, the model extends the Cubic Covariant Galileon (hereafter G3) \cite{Deffayet:2009wt} by including power laws of the field derivatives in the K-essence and cubic terms which still allow for tracker solutions. Contrary to the GCCG, the G3 model has been ruled out by cosmological data \cite{Renk:2017rzu,Peirone:2017vcq}.

While GCCG has been widely studied \cite{Giacomello:2018jfi,Frusciante:2019puu} both at background level and linear scales,  the non-linear, small scales are yet to be explored. The non-linear regime is where the situation
starts to become challenging for MG models due to the complexity of the equations to be solved which usually require extremely expensive computations,  as well as new phenomenology arising from the screening mechanisms \cite{Joyce:2014kja} that needs to be taken into account. The latter are mechanisms able to recover GR at Solar System scales where no signature of modifications of the gravitational force has been found (see Ref.~\cite{Will:2014kxa} for example). In the case of Galileon models it is the Vainshtein mechanism, which acts through the second derivative of the scalar field \cite{Vainshtein:1972sx}.
Future surveys such as \textit{Euclid} \footnote{\textit{Euclid}: \url{www.euclid-ec.org}}, SKAO \footnote{Square Kilometer Array Observatory: \url{https://www.skao.int}} and Rubin-LSST \footnote{ The Vera C. Rubin Observatory Legacy Survey of Space and Time: \url{https://www.lsst.org}},  will largely improve the constraints on deviations from GR at these scales. Therefore the physics in the non-linear regime of cosmic structure formation 
 requires accurate and efficient modeling of non-linearities and the development of appropriate codes to interpret the data. 

In this paper we study  the modelling of non-linearities in the GCCG model using the halo model reaction method \cite{Cataneo:2018cic} which is based on the halo model and perturbation theory, and is implemented in the \texttt{ReACT} code~\citep[\href{https://github.com/nebblu/ACTio-ReACTio}{\faicon{github}}]{Bose:2020wch,Bose:2022vwi}. In particular we derive the necessary theoretical quantities and provide the implementation in \texttt{ReACT}. We then use this machinery to forecast how well a spectroscopic survey of galaxies and an imaging survey from a \textit{Euclid}-like mission  and  SKAO-like one can be used to constrain the GCCG model.  

The paper is organized as follows: In Section \ref{Sec:GCCG} we review the dynamics of the background evolution and linear growth of perturbations of the GCCG.  We also present  a detailed study of the collapse of a spherical matter overdensity, the non-linear growth of large scale structures and the computation of the the non-linear power spectrum at 1-loop order in this model. In Section \ref{Sec:Reaction} we revisit the halo model reaction framework and in Section \ref{Sec:validation} we validate our approach against $N$-body simulations. In Section \ref{Sec:nonlinearphenomenology} we show the phenomenology of the GCCG model at non-linear scales and in Section \ref{Sec:Forecasts} we provide forecasts for \textit{Euclid}-like and SKO-like surveys. Finally we conclude in Section \ref{Sec:Conclusion}.

\section{Generalized Cubic Covariant Galileon}\label{Sec:GCCG}

We start by considering the cubic Horndeski theory described by the action~\cite{Deffayet:2010qz,Kobayashi:2010cm}
\ba\label{action1}
S=\int d^4x\sqrt{-g}\l(\f{M_{\rm pl}^2}{2} R +L_2+L_3\r)+ S_m[g_{\mu\nu},\chi_i], \nn \\  
\ea
where $M_{\rm pl}^2$ is the Planck mass, $R$ is the Ricci scalar, $g$ is the determinant of the metric $g_{\mu\nu}$, $S_m$ stands for the matter action for all matter fields, $\chi_i$,  and the Lagrangians $L_i$ are defined as follows
\begin{eqnarray}
{ L}_2=G_2(\phi, X), \quad { L}_3=  G_3(\phi, X)  \Box\phi,
\end{eqnarray}
with $G_i$ being free functions of the scalar field $\phi$ and $X=\partial_\mu \phi\partial^\mu\phi$. 

In this work we choose the forms of the $G_i$ functions as follows~\cite{DeFelice:2011bh}:
\ba
&&G_2=-c_2  \alpha _2^{4 \left(1-p_2\right)} \l(-X\r)^{p_2},\quad G_3=-c_3  \alpha _3^{1-4 p_3}  \l(-X\r)^{p_3}\,,\nn\\
&&
\ea
where $c_i, \alpha_i, p_i$ are constants with
\be
\alpha_2=\sqrt{H_0 M_{\rm pl}}\,,\quad \alpha_3=\left(\frac{M_{\rm pl}^{1-2 p_3}}{H_0^{2 p_3}}\right){}^{\frac{1}{1-4 p_3}}\,,
\ee
with $H_0$ being the Hubble constant. Without loss of generality we set
 $c_2=1/2$~\cite{Barreira:2013xea,2014JCAP...08..059B,Renk:2017rzu}. The above model is known to be the Generalized Cubic Covariant Galileon (hereafter GCCG) and it extends the original proposal of the Covariant Galileon~\cite{Deffayet:2009wt} (hereafter G3) because it considers  power law functions of $X$ in the Lagrangians, i.e. $G_i \propto X^{p_i}$. Indeed the G3 model is obtained by setting $p_2=p_3=1$.
 
 \subsection{Background evolution}

At background level if we consider the flat Friedmann-Lema\^{i}tre-Robertson-Walker (FLRW) metric 
\be
ds^2=-N(t)\,dt^2+a^2(t)\delta_{ij}dx^idx^j \,,
\ee
where $a(t)$ is the scale factor, $N(t)$ is the lapse function and $t$ is the cosmic time, we can vary \autoref{action1} with respect to $a$ and $N$ and we obtain
\ba
3M_{\rm pl}^2H^2&=&(\rho_{\rm m}+\rho_\phi)\,,\label{Fri1} \\
M_{\rm pl}^2(2\dot{H}+3H^2)&=& -(p_{\rm m}+p_\phi)\,,\label{Fri2}
\ea
where $H\equiv\dot{a}/a$ is the Hubble rate, an over-dot stands for derivatives with respect to $t$, $\rho_{\rm m}$ and $p_{\rm m}$ are the density and pressure of the standard matter fluids and 
\ba
&&\rho_\phi= 2XG_{2,X}-G_2-6aXH\phi^\prime G_{3,X}-XG_{3,\phi}\,,\\
&&p_\phi= G_2+2X\l(a(\dot{H}+H^2)\phi^\prime-a^2H^2\phi^{\prime\prime}\r)G_{3,X}-XG_{3,\phi}\,,\nn\\
&&
\ea
are the density and pressure associated to the scalar field.  Here the prime is the derivative with respect to the scale factor, $G_{i,X}=\partial G_i/\partial X$ and $G_{i,\phi}=\partial G_i/\partial \phi$.  For the matter fields, which we assume to be perfect fluids, we consider the continuity equation:
\be
\dot{\rho}_{\rm m}+3H(1+w)\rho_{\rm m}=0\,,
\ee
with $w\equiv p_{\rm m}/\rho_{\rm m}$. 
Finally, the equation of evolution for $\phi$ can be obtained by varying \autoref{action1} with respect to the scalar field itself. It reads
\be\label{phieq}
\f{H}{a^2}\f{\rm d}{{\rm d} a}\l(a^3 J\r)=P\,,
\ee
where 
\ba\label{eq:phi}
&&J=-2 aH \phi'G_{2,X}-6HX G_{3,X}+2aH \phi'G_{3,\phi}\,,\\
&&P=G_{2,\phi}-2X\bigg\{G_{3,\phi\phi}-2\bigg[a(\dot{H}+H^2)\phi^\prime \nn\\
&&\hspace{0.3cm} -a^2H^2\phi^{\prime\prime}\bigg]G_{3,\phi X}\bigg\}\,.
\ea

The GCCG model at the background level is characterized by a tracker solution which is given by~\cite{DeFelice:2011bh}
\be\label{tracker}
H^{2q+1}\psi^{2q}=\zeta H_0^{2q+1}\,, 
\ee
where we have defined a dimensionless constant
\ba
 q\equiv(p_3-p_2) +\frac{1}{2} \,,
\ea  
and a dimensionless scalar field:
\be
\psi\equiv\f{1}{M_{\rm pl}}\f{d \phi}{d\ln a}\,.
\ee
We have also introduced 
a dimensionless constant $\zeta$.  We can implement the above tracker solution in the background  \autoref{Fri1}, which simply reads \cite{Frusciante:2019puu}: 
\be\label{eq:background}
\l(\f{H}{H_0}\r)^{2+s}=\Omega_\phi^0+\l[\f{\Omega_c^0+\Omega_b^0}{a^3}+\f{\Omega_r^0}{a^4}\r]\l(\f{H}{H_0}\r)^{s}\,,
\ee
where $s=p_2/q$ and $\Omega_{i}^0\equiv \rho_{i}^0/3M_{\rm pl}^2H_0^2$ is the density parameter at present time for cold dark matter, $c$, baryonic matter, $b$, radiation, $r$, and scalar field, $\phi$. The latter  can be computed by evaluating the above equation at the present time, which gives the flatness condition:
\ba\label{flatness}
\Omega_\phi^0&=&1-\Omega_{\rm m}^0=c_3 (2 s\,q+2 q-1) \zeta ^{s+1}-\frac{1}{6} (2 s\,q -1)\zeta ^{s}\,,\nn\\
&&
\ea
where $\Omega_{\rm m}^0=\Omega_c^0+\Omega_b^0+\Omega_r^0$. It can be easily verified that:
\ba
\zeta=\l(6\Omega_\phi^0\r)^{\f{1}{s}}\,,\quad c_3=\f{1}{3}\f{s \, q }{\l(6\Omega_\phi^0\r)^{\f{1}{s}}(2s \, q+2q-1)}\,\,,
\ea
therefore the GCCG model has only two extra free parameters, i.e. $\{s, q\}$, with respect to $\Lambda$CDM.   In this description the G3 model does not have additional parameters because $s=2$ and $q=1/2$.  The GCCG model has been analysed using  data by \textit{Planck} and it has been found at 95\% C.L. that  $q>0$ and $s=0.6^{+1.7}_{-0.6}$ and a preference over $\Lambda$CDM has been found using the effective $\chi_{\rm eff}^2 $ and Deviance Information Criterion \cite{Frusciante:2019puu}. Including the measurements of baryon acoustic oscillation, SNIa and redshift space distortion is possible to find a lower bound for $q$, i.e. $q>0.8$, while $s=0.05^{+0.08}_{-0.05}$. We note that due to stability conditions (ghost and gradient requirements) both parameters are restricted to be positive. 

\subsection{Linear growth of perturbations}

At the linear perturbation level we can consider the perturbed FLRW metric in the Newtonian gauge \footnote{We note that the definition of the metric in Ref. \cite{Takushima:2015iha} uses an opposite convention: $\delta g_{00}=2\Phi$ and $\delta g_{ij}=-2\Psi\delta_{ij}$. For the model under consideration we will see that $\Phi=\Psi$.}:
\begin{equation}
 \mathrm{d}s^2 = -(1+2\Psi)\mathrm{d}t^2 + a^2(1-2\Phi)\delta_{ij}\mathrm{d}x^i\mathrm{d}x^j\,,
\end{equation}
where  $\Phi(t,x_i)$ and $\Psi(t,x_i)$ are the two gravitational potentials. We can write general forms for the linear perturbation equations in MG which relate the gravitational potentials and the matter perturbation $\delta_{\rm m}\equiv \delta\rho_{\rm m}/\rho_{\rm m}$, where $\rho_{\rm m}$ is the background matter density. They are~\cite{Amendola:2007rr,Bean:2010zq,Silvestri:2013ne,Pogosian:2010tj,Amendola:2019laa}:
\ba
\label{mudef}
&& -k^2\Psi = 4\pi G_{\rm N} a^2\mu_{\rm L}(a,k)\rho_{\rm m}\delta_{\rm m}\,, \\
&& -k^2(\Psi+\Phi) = 8\pi G_{\rm N} a^2\Sigma_{\rm L}(a,k)\rho_{\rm m} \delta_{\rm m}\,,\label{sigmadef}
\ea
where $k$ is the comoving wavenumber, $G_{\rm N}= (8\pi M_{\rm pl}^{2})^{-1}$ is  the Newtonian gravitational constant, $\mu_L$ and $\Sigma_L$ are respectively the linear effective gravitational coupling and light deflection parameter. For the GCCG model, when using the  Quasi Static Approximation (QSA)\footnote{The QSA assumes that the time derivatives of the perturbed quantities can be neglected when compared with their spatial derivatives. This approximation is valid for modes deep inside the Hubble radius \cite{Sawicki:2015zya}.} they read:
\begin{eqnarray}
  \mu_L(a)=  \left(1+\dfrac{2\alpha_B^2}{c_s^2\alpha} \right)\;, \qquad \Sigma_L(a)=\mu_L\;,
\end{eqnarray}
with $\alpha \equiv \ak +6 \ab^2>0$ being the no-ghost condition with 
\ba
 \ak =  -12q \, \ab\,,  \qquad
 \ab = - s \, q \, \Omega_{\phi }^0  \l(\frac{H_0}{H}\r)^{s+2}\,,
\ea
being respectively the kineticity and the braiding functions. Finally, $c_s^2$ is the speed of propagation of the scalar mode and it is defined as
\begin{equation}\label{eqn:cs2}
 c_{\rm s}^2 =\frac{2}{\alpha}\l[ (1+\alpha_{\rm B})\l(-\frac{\dot{H}}{H^2}-\ab\r)
 -\frac{\dot{\ab}}{H}\r]
 -\frac{3\Omega_{\rm m}}{\alpha}\,.
\end{equation}
The latter has to be positive to avoid gradient instability. For a thorough discussion about the viability space of the GCCG model we forward the reader to Refs. \cite{Giacomello:2018jfi,Frusciante:2019puu} 
 
Finally we have to consider the  evolution equation of the matter density perturbation which reads:
\be\label{eqn:linpert}
\frac{{\rm d}^2\delta_{\rm m}}{{\rm d}^2 \ln a} + \l(2+\f{1}{H}\frac{{\rm d} H}{{\rm d} \ln a}\r) \frac{{\rm d} \delta_{\rm m}}{ {\rm d} \ln a} - \f{3}{2}\Omega_{\rm m} \mu_{\rm L}(a) \delta_{\rm m}=0\,,
\ee
where $\Omega_{\rm m}(a)=\rho_{\rm m}/3M_{\rm pl}^2H^2$ is the time dependent matter density parameter.

\subsection{Non-linear growth of perturbations}\label{Sec:nonlinear}

Let us consider non-linear perturbations and assume the  QSA holds. We will  consider  for modelling the  formation of gravitationally bound structures, the spherical collapse model to describe the collapse of what we assume to be a spherical overdensity. The collapse phase  is then followed by a phase where  virialization takes place.   The methodology adopted here has been already applied to MG models \cite{Schmidt:2009yj,Kimura:2010di,Bellini:2012qn,Barreira:2013eea,Frusciante:2020zfs,Albuquerque:2024hwv}.

The non-linear equations for the GCCG model read \cite{Albuquerque:2024hwv}:
\ba
&&\nabla^2\Psi=\frac{\rho \,  \delta_{\rm m}}{2M_{\rm pl}^2} +\ab \nabla^2Q \,, \label{PoissonNL}\\
&&\Phi=\Psi\,,\\
&&\nabla^2Q+\lambda^2\left[(\nabla_i\nabla_jQ)^2 - \left(\nabla^2Q\right)^2\right]=-\frac{\lambda^2H^2}{2M_{\rm pl}^2}\rho \, \delta_{\rm m }\,, \nn \\  \label{secondfield}
\ea
where we have defined: 
\ba
&&Q=\frac{H\delta\phi}{\dot{\phi}}\,, \qquad \lambda^2=-\f{ 2 \ab}{H\alpha c_s^2}\,.
\ea

Combining the above equations, assuming a top-hat profile for the density field and after some integrations, we found a modified Poisson equation  which includes non-linear corrections:
\be\label{nonlinearpoisson}
\f{\partial^2 \Psi}{a^2} = 4\pi G_{\rm N} \mu^{\rm NL}(a,R)\,
\rho_{\rm m}\delta_{\rm m}\,,
\ee
where we define $\mu^{\rm NL}$ as the non-linear effective gravitational coupling and its form is 
\ba\label{muNL}
\mu^{\rm NL}(a,R) = 
1 + 2\l(\mu^{\rm L}-1\r)\l(\f{R}{R_{\rm V}}\r)^3
\l(\sqrt{1+\f{R_{\rm V}^3}{R^3}}-1\r)\,,\, \nn\\
&&
\ea
with 
\be
\left(\frac{R_V}{R}\right)^3=4H_0^2\Omega_{\rm m}^0\lambda^4 \frac{\delta_{\rm m}}{a^3} \,,
\ee
where $R_{\rm V}^3=8 G_{\rm N}\lambda^4 \delta M$\, is the the Vainshtein radius, $\delta M$ being the total mass of the density perturbation $\delta \rho_{\rm m}$ and $R$ is the radius of the sphere of mass $M$.  The Vainshtein mechanism plays a very important role when considering the formation of gravitationally bound structures. Indeed it suppresses the modification to the gravitational force in high-density environments which is what happens  during the collapse phase when the density of the region is sufficiently high to significantly modify the dynamics of the scalar field.

\subsubsection{Spherical collapse}\label{Sec:Sphericalcollapse}

The general form  of the  non-linear evolution equation for the matter density is:
\be
\ddot{\delta}_{\rm m} + 2H\dot{\delta}_{\rm m} -  \f{4}{3}\f{\dot{\delta}_{\rm m}^2}{1+\delta_{\rm m}} = \l(1+\delta_{\rm m}\r)\f{\partial^2\Psi}{a^2}\,,
\ee
which for MG assumes the following form when the Poisson equation is used
 \be\label{nonlinearoverdensityeq}
\ddot{\delta}_{\rm m} + 2H\dot{\delta}_{\rm m} - \f{4}{3}\f{\dot{\delta}_{\rm m}^2}{1+\delta_{\rm m}} = 4\pi G_{\rm N} \mu^{\rm NL}\,
\rho_{\rm m}\l(1+\delta_{\rm m}\r)\delta_{\rm m}\,.
\ee
The above equation can be used to obtain the equation for the radius of the spherical top-hat, $R$, by starting from the assumption that 
the total mass inside $R$ is conserved during the collapse phase, i.e.  
\be \label{eq:M}
M = \f{4\pi}{3}R^3\rho_{\rm m}(1+\delta_{\rm m})=\text{const} \,.
\ee
By performing a change of variable
\be\label{defy}
 y=\f{R}{R_i} - \f{a}{a_{i}}\,,
\ee
where $R_{i}$ and $a_{i}$  are respectively the initial values of $R$ and  scale factor, and  differentiating \autoref{eq:M}, one finds  
\ba\label{yeq}
\frac{{\rm d}^2y}{{\rm d} \ln a^2}& =& -\f{1}{H}\frac{{\rm d} H}{{\rm d}\ln a}\frac{{\rm d} y}{{\rm d}\ln a} + \l(1+\f{1}{H}\frac{{\rm d} H}{{\rm d} \ln a}\r)y \nn\\&-& \f{\Omega_{\rm m}}{2}\mu^{\rm NL}\delta_{\rm m}\l(y +\f{a}{a_{i}}\r)\,.
\ea
 Finally, we can write the overdensity as
\be
 \delta_{\rm m} = (1+\delta_{{\rm m},i})\l(1+\f{a_{i}}{a}y\r)^{-3}-1\,,
\ee
which follows from matter conservation.

\subsubsection{Virial Theorem}\label{Sec:virialtheorem}

Finally we consider the last stage of the collapse: the virialization, i.e. when the collapse stops and the system reaches equilibrium,  and satisfies the Virial theorem.
The Virial Theorem reads
\be
W+2T=0\,,
\ee
where $W$ is the potential energy  and $T$ the kinetic energy. The kinetic energy has the form
\be
\frac{T}{E_0}=\frac{H^2}{H_0^2}\l[\frac{a}{a_i}\l(\frac{{\rm d}y}{{\rm d}\ln a}+y\r)\r]^2\,,
\ee
with $E_0=3/10M(H_0R_i)^2$.
The potential energy gets three energy contributions: the Newtonian ($W_N$), scalar field ($W_\phi$) and background ($W_{\rm eff}$) potential energies, which read \cite{Schmidt:2009yj,Cataneo:2018cic}
\ba
&&\frac{W_N}{E_0} = -\Omega_{\rm m}^0 y^2 (1+\delta_{\rm m})\frac{a^{-1}}{a_i^2}\,,\\
&&\frac{W_\phi}{E_0}=-\Omega_{\rm m}^0\mathcal{F}y^2\delta_{\rm m} \frac{a^{-1}}{a_i^2}\,,\\
&& \frac{W_{\rm eff}}{E_0}= -\frac{8\pi G_N}{3H^2_0}(1+3 w_{\rm eff})\bar{\rho}_{\rm eff} \frac{a^2}{a^2_i}y^2\,,
\ea
where for the GCCG we have defined $\mathcal{F}\equiv\mu^{NL}-1$ in \autoref{muNL} and
\ba
&& \bar{\rho}_{\rm eff}=\left(\frac{H_0}{H}\right)^s \frac{3H_0}{8\pi G_N}\Omega_\phi^0\,,\\
&&w_{\rm eff}=\frac{\dot{H}s}{3H^2}-1\,,
\ea
according to which 
\be
\frac{W_{\rm eff}}{E_0}=\left(2-\frac{\dot{H}}{H^2}s\right)\left(\frac{H}{H_0}\right)^2\left(1-\Omega_{\rm m}\right)\frac{a^2}{a^2_i}y^2\,.
\ee

\subsubsection{Higher order coupling kernels of the power spectrum at 1-loop order} \label{Sec:kernels}

In this section we compute the non-linear power spectrum at 1-loop order using standard perturbation theory (SPT) \cite{bernardeau_large-scale_2001}. We consider the relation between $\Psi$ and $\delta_{\rm m}$ and expand up to 3rd order in the matter perturbation \cite{Bernardeau:2001qr,Koyama:2009me}
\be
-\frac{k^2}{a^2H^2}\Psi=\frac{3}{2}\Omega_{\rm m}\mu_L(k,a)\delta_{\rm m}({\bf k}) +S({\bf k})\,,
\ee
where $S({\bf k})$ is the non-linear source term up to the third order and it is given by \citep{Bose:2016qun}
\begin{widetext}
\ba
S({\bf k})&=& \int\frac{d^3{\bf k}_1d^3{\bf k}_2}{(2\pi)^3}\delta_D({\bf k}-{\bf k}_{12})\gamma_2({\bf k},{\bf k}_1,{\bf k}_2;a)\delta_{\rm m}({\bf k}_1)\delta_{\rm m}({\bf k}_2)\nn\\
&+&\int\frac{d^3{\bf k}_1d^3{\bf k}_2d^3{\bf k}_3}{(2\pi)^6}\delta_D({\bf k}-{\bf k}_{123})\gamma_3({\bf k},{\bf k}_1,{\bf k}_2,{\bf k}_3;a)\delta_{\rm m}({\bf k}_1)\delta_{\rm m}({\bf k}_2)\delta_{\rm m}({\bf k}_3)\,,
\ea
\end{widetext}
where $\delta_D$ is the Dirac delta function,  ${\bf k}_{ij}={\bf k}_i+{\bf k}_j$ and  ${\bf k}_{ijk}={\bf k}_1+{\bf k}_j+{\bf k}_k$, $\gamma_2({\bf k},{\bf k}_1,{\bf k}_2;a)$ and $\gamma_3({\bf k},{\bf k}_1,{\bf k}_2,{\bf k}_3;a)$ are functions symmetric  under the exchange of ${\bf k}_i$. We follow Ref.~\cite{Bose:2016qun} to compute their forms for the GCCG and we find:
\begin{widetext}
\ba
&&\gamma_2({\bf k},{\bf k}_1,{\bf k}_2;a)=-18\left(\frac{H_0}{H}\right)^4\frac{ \ab^4}{(\alpha c_s^2)^3} \left(\frac{\Omega_{\rm m}^0}{a^{3}}\right)^2\left(1-\frac{({\bf k}_1 \cdot {\bf k}_2)^2}{k_1^2 k_2^2}\right)\,,\\
&&\gamma_3({\bf k},{\bf k}_1,{\bf k}_2,{\bf k}_3;a)=72\left(\frac{H_0}{H}\right)^6\frac{ \ab^6}{(\alpha c_s^2)^5} \left(\frac{\Omega_{\rm m}^0}{a^{3}}\right)^3\left(1-\frac{({\bf k}_1 \cdot {\bf k}_{23})^2}{k_1^2 k_{23}^2}\right)\left(1-\frac{({\bf k}_2 \cdot {\bf k}_3)^2}{k_2^2 k_3^2}\right)\,.
\ea
\end{widetext}

\section{The halo model Reaction}\label{Sec:Reaction}

In this Section we review a model-independent framework to compute the non-linear matter power spectrum, namely the halo model reaction \cite{Cataneo:2018cic}.  This approach combines both 1-loop perturbation theory and the halo model.

The non-linear matter power spectrum of a modified theory of gravity, hereafter $P_{\rm NL}$, is computed as follows
\be
P_{\rm NL}=\mathcal{R}(k,z)P_{\rm NL}^{\rm pseudo}(k,z)\,,
\label{eq:nlpk}
\ee
where $P_{\rm NL}^{\rm pseudo}$ is the non-linear pseudo power spectrum. This is a the non-linear  matter power spectrum within $\Lambda$CDM whose initial conditions are such that the modified linear clustering is reproduced at the target redshift. The second component, $\mathcal{R}$, is the halo model reaction given by
\begin{align}
&\mathcal{R}(k,z) = \nonumber \\
&\frac{\{[1-\mathcal{E}(z)]e^{-k/k_\star(z)} + \mathcal{E}(z)\} P_{\rm L}(k,z)  +  P_{\rm 1h}(k,z)}{P_{\rm hm}^{\rm pseudo}(k,z)}\,,
\end{align}
with $P_{\rm L}$ being the MG linear matter power spectrum and $P_{\rm 1h}$ being the 1-halo contribution to the power spectrum as predicted by the halo model for the MG scenario (see Section \ref{Sec:Sphericalcollapse} for the GCCG case). Finally we have:
\begin{align}
  P_{\rm hm}^{\rm pseudo}(k,z) = &   P_{\rm L} (k,z) + P_{\rm 1h}^{\rm pseudo}(k,z), \label{Pk-halos} \\ 
  \mathcal{E}(z) =& \lim_{k\rightarrow 0} \frac{ P_{\rm 1h}^{\rm }(k,z)}{ P_{\rm 1h}^{\rm pseudo}(k,z)} , \label{mathcale} \\ 
   k_{\rm \star}(z) = & - \bar{k} \left\{\ln \left[ 
    \frac{A(\bar{k},z)}{P_{\rm L}(\bar{k},z)} - \mathcal{E}(z) \right] \right.\nn\\
    &\left.- \ln\left[1-\mathcal{E}(z) \right]\right\}^{-1}\,, \label{kstar}
\end{align}
where $\bar{k}=0.06$ h~Mpc$^{-1}$ and the $k \rightarrow 0$ limit in the above is taken to be $k=0.01$ h~Mpc$^{-1}$ according to Ref.~\cite{Cataneo:2018cic} and
\ba
    A(k,z) &=&  \frac{P_{\rm 1-loop}(k,z)+ P_{\rm 1h}(k,z)}
    {P^{\rm pseudo}_{\rm 1-loop}(k,z)+ P_{\rm 1h}^{\rm pseudo}(k,z)}  P_{\rm hm}^{\rm pseudo}(k,z) \nn\\
    &-&  P_{\rm 1h}(k,z)\,.
\ea
In the above expressions we further define $P_{\rm 1h}^{\rm pseudo}$ as the  1-halo contribution to the power spectrum as predicted by the halo model for the pseudo cosmology, $P_{\rm 1-loop}$ and $P^{\rm pseudo}_{\rm 1-loop}$ are respectively the 1-loop matter power spectra from SPT with and without non-linear MG effects ($S({\bf k}) =0$). The higher order coupling kernels of $P_{\rm 1-loop}$ for the GCCG model are calculated in Section \ref{Sec:kernels}. 

We modify the public software package \texttt{ReACT}  ~\citep[\href{https://github.com/nebblu/ACTio-ReACTio}{\faicon{github}}]{Bose:2020wch,Bose:2022vwi}  which allows for 
fast and accurate calculation of non-linear power spectra for beyond standard models using the reaction method reviewed above. Our new patch
accounts for a modified background evolution (\autoref{eq:background}), the spherical collapse and virial theorem computations
defined respectively in Secs. \ref{Sec:Sphericalcollapse} and \ref{Sec:virialtheorem}, which are needed for the halo-model
spectra computations, and the 1-loop corrections  in Sec. \ref{Sec:kernels} to compute the 1-loop matter power spectra. Note our implementation also allows the computation of the redshift space power spectrum multipoles in GCCG \citep{Bose:2016qun}. 

We have compared the background evolution and the linear matter power spectrum of the GCCG of our \texttt{ReACT} patch with the Einstein-Boltzmann solver \texttt{EFTCAMB} \cite{Raveri:2014cka,Hu:2013twa} where the GCCG is implemented \cite{Frusciante:2019puu}, and we find agreement at the sub-percent level. For the non-linear matter power spectrum implementation we discuss its validation in the next section.

Note that for the non-linear spectrum, we must also choose a prescription for $P_{\rm NL}^{\rm pseudo}$ in \autoref{eq:nlpk}. For this we make two choices in the draft: halofit \citep{takahashi_revising_2012} and \texttt{HMCode2020} \citep{Mead:2020vgs}. We would normally expect \texttt{HMCode2020} to be the most accurate prescription, but what we have found is that this prescription gives poor results when comparing \autoref{eq:nlpk} directly with the simulation measurements. This is likely due to the exceptionally high $\sigma_8(z=0)$ of the simulations, which is far beyond the values to which \texttt{HMCode2020} was fit to. Further, it was found  that the \texttt{HMCode2020}-based non-linear boost factor, $B(k,z) = \mathcal{R}(k,z) \times P_{\rm HMCode2020}^{\rm pseudo}(k,z) /P_{\rm HMCode2020}^{\rm \Lambda CDM} \leq 1$ at quasi-non-linear scales. This is not to be expected from such scalar tensor theories which act to enhance power. This effect was not found for \texttt{HMCode2016} nor halofit. 

In particular, \texttt{HMCode2020} introduces a $\sigma_8$ dependent damping to the linear power spectrum which can cause exactly this effect, as well as over damping of the power spectrum if $\sigma_8$ is exceptionally high, as in the simulations considered in this work. For this reason, we choose to use halofit when comparing to the simulations which gives the level of agreement seen in many other works \citep{Bose:2022vwi,Cataneo:2018cic,Bose:2020wch,Bose:2021mkz,Carrilho:2021rqo,Parimbelli:2022pmr,Euclid:2022qde,Srinivasan:2023qsu,Srinivasan:2021gib}. In our forecasts, we switch to \texttt{HMCode2020} as the fiducial $\sigma_8$ is chosen to be much lower, where the \texttt{HMCode2020} prescription for the pseudo spectrum should outperform halofit. In an upcoming work we demonstrate this improvement explicitly.

\section{Validation of the modelling against simulations}\label{Sec:validation}


\begin{figure*}[t!]
    \centering
\includegraphics[width=0.95\textwidth]{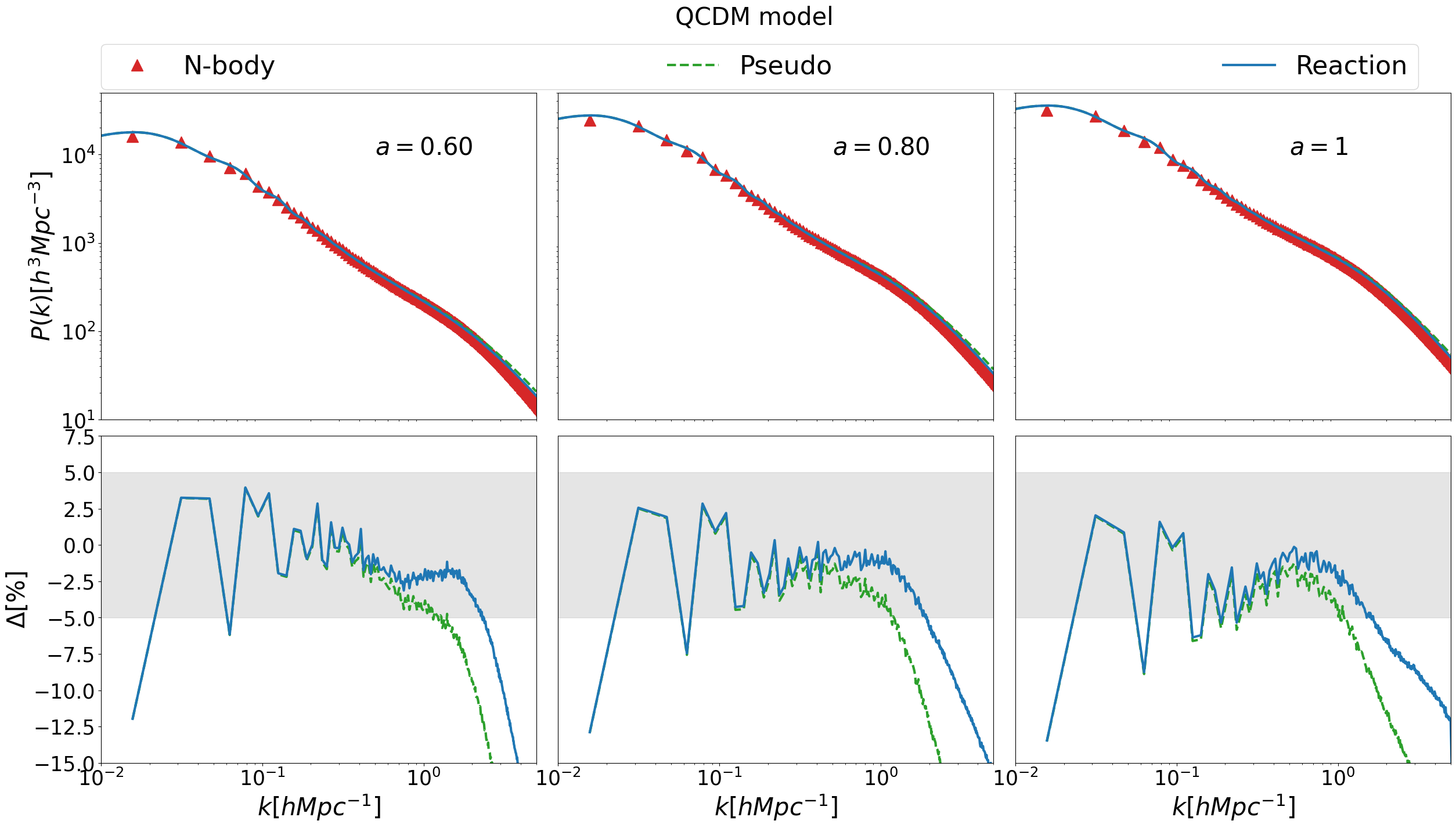}
    \caption{Top panels: Non-linear matter power spectra for the QCDM model at three times: $a=0.60, 0.80$ and $1$.  Red triangles are measurements from the $N$-body simulations,  green dashed lines are the pseudo spectrum and blue solid lines are the halo model reaction predictions. Bottom panels: Relative percentage difference of the model predictions  vs $N$-body simulations for the non-linear matter power spectra, $\Delta=100\% (1-P_{\rm prediction}/P_{\rm N-body})$.}
    \label{fig:QCDMvalidation}
\end{figure*}
\begin{figure*}[t!]
\includegraphics[width=0.95\textwidth]{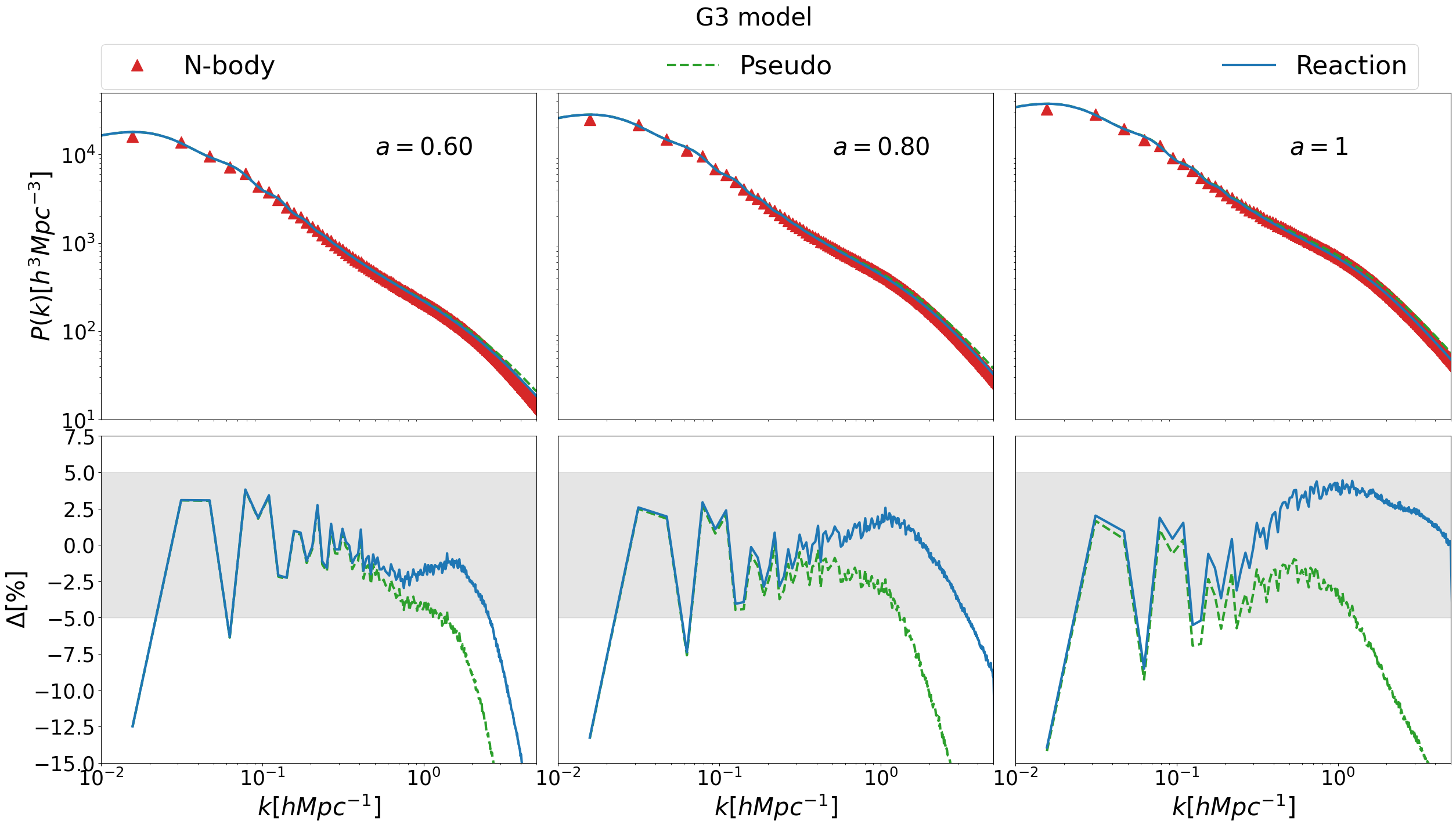}
    \caption{Top panels: Non-linear matter power spectra for the G3 model at three times: $a=0.60, 0.80$ and $1$.  Red triangles are measurements from the $N$-body simulations,  green dashed lines are the pseudo spectrum and blue solid lines are the halo model reaction predictions. Bottom panels: Relative percentage difference of the model predictions  vs $N$-body simulations for the non-linear matter power spectra, $\Delta=100\% (1-P_{\rm prediction}/P_{\rm N-body})$.}
    \label{fig:G3validation}
\end{figure*}


$N$-body cosmological simulations of the GCCG do not exist yet but in the limit $s=2$ and $q=1/2$ the model reduces to the G3 model for which  simulations were run using the \texttt{ECOSMOG} code \cite{Li:2011vk,Barreira:2013eea,Becker:2020azq}. We will employ these simulations to validate our implementation of the non-linear matter power spectrum in the G3 limit. The simulations are dark matter only and the box sizes are $L=200$ Mpc\,$h^{-1}$ and $L=400$ Mpc\,$h^{-1}$ with threshold values $N_{\rm p,th}=8$ and $N_{\rm p,th}=6$ respectively. In both cases the total number of particles is $N_{\rm p}=512^3$ and the domain grid has 512 cells in each direction. The reference cosmology for these simulations considers the best fits obtained by running Monte Carlo
Markov chains and combining WMAP9+SNLS+BAO data \cite{Barreira:2013jma}:  $\Omega_{\rm r}^0h^2=4.28\times10^{-5}$, $\Omega_{\rm b}^0h^2=0.02196$, $\Omega_{\rm c}^0h^2=0.1274$, $h=H_0/100=0.7307$, the primordial spectra index $n_{\rm s}=0.953$, the optical depth to reionization $\tau=0.0763$, the scalar amplitude $A_{\rm s}$ at pivot scale $k=0.02$ Mpc$^{-1}$ is $\log [10^{10}A_{\rm s}]=3.154$. The background is solved differently than our approach, where $c/c_3^{2/3}=-5.378$ and $c_3=10$ and $\log[\rho_{\phi,i}/\rho_{m,i}]=-4.22$, where ``i'' refers to quantities evaluated at initial time, $z_i=10^6$.   We adopt the same values for the cosmological parameters for comparison purposes but note that $\Omega_{\rm r}^0=0 $ is assumed in \texttt{ReACT}. 

Besides the simulations for the G3 model, in the same work the authors run simulations for another model, 
 QCDM, which is defined such that the background is the same as G3 and the perturbations are those of $\Lambda$CDM. This was made with the purpose of disentangling the effects of the modified gravitational strength in changing the linear matter growth from those of the background. We have also implemented this scenario in our \texttt{ReACT} patch.

In \autoref{fig:QCDMvalidation} and \autoref{fig:G3validation}  we present  the results of the comparison for QCDM  and the G3   models respectively.  We show results for both the pseudo spectra and full halo model reaction predictions along with the matter power spectra measured from the $N$-body simulations for three different times ($a=0.60, 0.80, 1$ or $z=0.67, 0.25, 0$) and for box size $L=400$ Mpc\,$h^{-1}$. From \autoref{fig:QCDMvalidation} we can infer differences coming from the background evolution only. Let us stress that the $N$-body simulations and the theoretical predictions use different approaches to solve the background evolution which can lead to small differences. Despite this, for  all three scale factors, we have agreement between theory and simulation within 5\%  for $2 \times 10^{-2}~h\,{\rm Mpc}^{-1}\lesssim k\lesssim 1$ $h$\,Mpc$^{-1}$. Further, for $0.2 ~h\,{\rm Mpc}^{-1}\lesssim k\lesssim 0.7$ $h$\,Mpc$^{-1}$ the agreement is better (within 3\%). 
For the G3 model the accuracy is the same as for QCDM (see \autoref{fig:G3validation}).  For the reaction predictions it is even possible to find agreement within the 5\% region for  smaller scales ($k\lesssim 3$ $h$\,Mpc$^{-1}$ at $a=0.6$ and $a=0.8$ and $k\lesssim 5$ $h$\,Mpc$^{-1}$ at $a=1$).

Recall that the accuracy of our pseudo spectrum, halofit, is $\sim 5\%$ at these scales \citep{takahashi_revising_2012}, implying improvement in the pseudo can lead to far better predictions. Slightly better accuracy ($\sim 2-3\%$) was found for many other non-standard models within the halo model reaction formalism \citep{Cataneo:2018cic,Cataneo:2019fjp,Bose:2021mkz,Srinivasan:2021gib,Carrilho:2021rqo,Parimbelli:2022pmr,Bose:2022vwi,Euclid:2022qde} where the authors typically compare the theoretical predictions to simulation measurements of the ratio of power spectra with respect to $\Lambda$CDM, which are not available in this case. The ratio typically factors out some of the systematic inaccuracy of the pseudo. 

We thus conclude that our predictions show the expected accuracy of the halo model reaction formalism as presented in the literature, which we deem sufficient up to scales $k\lesssim 1$ $h$\,Mpc$^{-1}$ for a wide redshift range, and that the reaction formalism can then be used to model the non-linearities of the Galileon model. We then extend the validity of the modelling to the GCCG as well.

\section{Phenomenology of the non-linear matter power spectrum for Generalized Cubic Covariant Galileon}\label{Sec:nonlinearphenomenology}
\begin{figure*}[t!]
    
\includegraphics[width=0.7\textwidth]{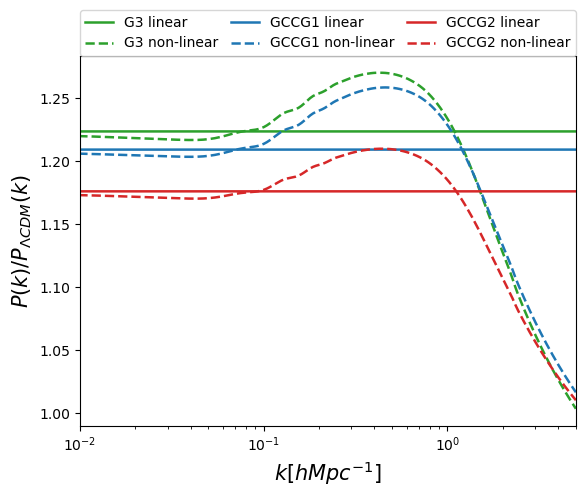}
    \caption{Ratio of the linear (solid lines) and non-linear (dashed lines) matter power spectra for the Galileon models with respect to their $\Lambda$CDM counterparts.}
    \label{fig:GCCGpheno}
\end{figure*}

In this Section we illustrate the phenomenology  of the GCCG by considering the non-linear matter power spectrum predictions from \texttt{ReACT}. In order to quantify the deviation with respect to the standard scenario and to the linear behaviour, we consider two sets of values for $\{s,q\}$:  $\{s=2,q=0.35\}$, we refer to it as GCCG1 and $\{s=1.3,q=0.5\}$, hereafter GCCG2. For comparison we include also G3 ($\{s=2,q=0.5\}$). For the cosmological parameters we use the same as in previous Section. We  show in 
\autoref{fig:GCCGpheno}  the ratio of the linear (solid lines) and non-linear (dashed lines) matter power spectra for  the Galileon models with respect to their  $\Lambda$CDM counterparts  at $z = 0$.
We note at linear scales an enhancement of the growth of structure with respect to $\Lambda$CDM for $ 0.01~h\,{\rm Mpc}^{-1}\lesssim k\lesssim 0.1$ $h\,{\rm Mpc}^{-1}$.  The enhancement is larger for the models with the higher values of $s$, i.e., G3 (green solid lines) and GCCG1 (blue solid lines). GCCG1 power spectra are lower than G3 ones because the value of $q$ is smaller. We can also notice that up to $k\lesssim 0.1 $ $h$ Mpc$^{-1}$  there is good agreement between the non-linear and the linear theory prediction ($\lesssim 0.5$\%). We can then deduce that the screening mechanism does not affect the large scales where usually the validity of the linear perturbation theory is assumed.
Non-linear corrections enter at smaller scales ($k\gtrsim 0.1$ $h$ Mpc$^{-1}$). We observe up to a $\sim 4\%$ enhancement  for $ 0.1~h\,{\rm Mpc}^{-1}\lesssim k\lesssim 1$ $h$ Mpc$^{-1}$ with respect to linear ratio and finally for the higher $k$ a suppression, showing the efficiency of the screening mechanism to restore standard gravity.

\section{Forecasts analysis for spectroscopic and photometric primary probes}\label{Sec:Forecasts}

We will provide  forecasts for the GCCG model using  spectroscopic and photometric galaxy clustering  (GC$_{\rm sp}$ and GC$_{\rm ph}$) and weak lensing (WL) probes. We will also include the cross correlation of GC$_{\rm ph}$ and WL (we will refer to this combination as XC) as well as the full combination of probes. We will use specifications of surveys such as \textit{Euclid}-like and SKAO-like.

\subsection{Method}
We follow the methodology adopted in Ref.~\cite{Euclid:2023rjj}, which extends the treatment of the spectroscopic and photometric probes  for $\Lambda$CDM described in Ref.~\cite{Euclid:2019clj} to scale-independent MG models. We summarize it here.

For the GC$_{\rm sp}$ probe we consider the observed galaxy power spectrum defined as:
\begin{multline}
P_\text{obs}(k, \mu_\theta ;z) = 
\frac{1}{q_\perp^2(z)\, q_\parallel(z)} 
\left\{\frac{\left[b\sigma_8(z)+f\sigma_8(z)\mu_{\theta}^2\right]^2}{1+ \Big[ f(z) k \mu_{\theta} \sigma_{\rm p}(z) \Big]^2 } \right\} 
\\
\times \frac{P_\text{dw}(k,\mu_{\theta};z)}{\sigma_8^2(z)}  
F_z(k,\mu_{\theta};z) 
+ P_\text{s}(z) \,, 
\label{eq:GC:pk-ext}
\end{multline}
where 
\begin{align}
q_{\perp}(z) &= \frac{D_{\rm A}(z)}{D_{\rm A,\, ref}(z)},\\
q_{\parallel}(z) &= \frac{H_\text{ref}(z)}{H(z)}\,,
\end{align}
accounts for the Alcock-Paczynski effect, defined in term of the angular diameter distance $D_A(z)$ and the Hubble parameter $H(z)$. The subscript ``ref'' stands for  reference cosmology. The term in  the curly bracket accounts for the redshift-space distortions where we define the effective scale-independent galaxy bias $b$, for which we consider numerical values  as in Ref.~\cite{Euclid:2019clj} (in the analysis we consider $\ln(b\sigma_8)$  as a free parameter  and we marginalize over), the growth rate, $f$, and  the square of the cosine of the angle between the wavevector $\bm k$ and the line-of-sight direction, $\mu_{\theta}^2$. We note that the denominator includes  the finger-of-God effect. We  consider $\sigma_{\rm v}^2 = \sigma_{\rm p}^2$, with 
\be
\sigma_{\rm v}^2 (z) = \frac{1}{6\pi^2} \int \de k P_{\rm L} (k,z) \,,\label{eq:sigmav}
\ee
and $P_{\rm L}$ being the linear matter power spectrum. They are evaluated  at every redshift bin, and we kept them fixed. Additionally $P_{\rm dw}$ defines the de-wiggled power spectrum, which includes the smearing of the baryon acoustic oscillations, which reads
\begin{equation}
P_\text{dw}(k,\mu_{\theta};z) = P_{\rm L}(k;z)\,\text{e}^{-g_\mu k^2} + P_\text{nw}(k;z)\left(1-\text{e}^{-g_\mu k^2}\right) \,,
\label{eq:pk_dw}
\end{equation}
where $P_{\rm nw}$ stands for a no-wiggle power spectrum. Non-linearities  are included in the  non-linear damping factor~\citep{Eisenstein:2006nj}
\be
g_\mu(k,\mu_\theta,z) = \sigma_{\rm v}^2(z) \left\{ 1 - \mu_\theta^2 + \mu_\theta^2 [1 + f(z)]^2  \right\}\,.
\ee
The function $F_z$  accounts for the redshift uncertainty as it is defined as
\begin{equation}
    F_z(k, \mu_{\theta};z) = \text{e}^{-k^2\mu_{\theta}^2\sigma_{r}^2(z)}\,,
\end{equation}
with $\sigma_{r}^2(z) = (1+z)\sigma_{z}/H(z)$, and $P_{\rm s}$ takes into account a residual shot noise, and it is considered as a nuisance parameter.

For the photometric probes we consider the  angular power spectra, which, under the Limber approximation, reads
\begin{equation}
 C^{XY}_{ij}(\ell) = \int_{z_{\rm min}}^{z_{\rm max}}\de z\,{\frac{W_i^X(z)W_j^Y(z)}{H(z)r^2(z)}P_{\rm NL}(k_\ell,z)}\,, \label{eq:ISTrecipe}
\end{equation}
where $i$ and $j$ stand for two tomographic bins, and $X$ and $Y$ represent either CG$_{\rm ph}$ or WL, 
$k_\ell=(\ell+1/2)/r(z)$, and $r(z)$ is the comoving distance. $P_{\rm NL}$ stands for the non-linear matter power spectrum, which we take from \texttt{ReACT} (see \autoref{eq:nlpk}).
Additionally we define the kernels for galaxy clustering and weak lensing,
\begin{widetext}
\begin{align}
 W_i^{\rm G}(k,z) =&\; b_i(k,z)\,\frac{n_i(z)}{\bar{n}_i}H(z)\,, \label{eq:wg_mg}\\  
 W_i^{\rm L}(k,z) =&\; \frac{3}{2}\Omegam \,H_0^2\,(1+z)\,r(z)\,\Sigma(z)
\int_z^{z_{\rm max}}{\de z'\frac{n_i(z')}{\bar{n}_i}\frac{r(z'-z)}{r(z')}}\nonumber\\
  &+W^{\rm IA}_i(k,z)\,. \label{eq:wl_mg}
\end{align}
\end{widetext}
In these expressions $n_i/\bar{n}_i$ is the normalised number density in the $i-$th bin.  The galaxy distribution is binned into 10 equi-populated redshift bins with a true distribution $n(z)\propto (z/z_0)^2\,{\rm exp}[-(z/z_0)^\gamma]$, with $z_0=z_m/\sqrt{2}$  being the median redshift, we will specify these numbers in the survey specifications. To account for the photometric redshift uncertainties  the redshift distribution is then convolved with a sum of two Gaussian
distributions. For details about the expressions see Ref.~\cite{Euclid:2019clj}.
Additionally, $b_i$  are the constant values of bias in each bin  and are  considered as nuisance parameters, then we marginalise over.  Their fiducial values are computed as $b_i(z)=\sqrt{1+\bar{z}}$ with $\bar{z}$ being the mean redshift of the bin (in this we follow Ref.~\cite{Euclid:2019clj}).
$\Sigma$ encodes the changes to the lensing potential for MG. In our case $\Sigma=\mu$. Finally $W^{\rm IA}_i$ defines the intrinsic alignment of galaxies with the extended nonlinear alignment model \cite{Euclid:2019clj}:
\begin{equation}\label{eq:IA}
 W^{\rm IA}_i(k,z)=-\frac{\mathcal{A}_{\rm IA}\,\mathcal{C}_{\rm IA}\,\Omega_{\rm m,0}\,\mathcal{F}_{\rm IA}(z)}{\delta(k,z)/\delta(k,z=0)}\frac{n_i(z)}{\bar{n}_i(z)}\,H(z)\,,
\end{equation}
and
\begin{equation}
 \mathcal{F}_{\rm IA}(z)=(1+z)^{\eta_{\rm IA}}\left[\frac{\langle L\rangle(z)}{L_\star(z)}\right]^{\beta_{\rm IA}}\,.
\end{equation}
with $\langle L\rangle(z)$ and $L_*(z)$  being the mean and the characteristic luminosity of source galaxies.


We will use specifications of surveys such as \textit{Euclid}-like and SKAO-like,  which are very close to the more realistic cases. In detail: 
\begin{itemize}
\item for a \textit{Euclid}-like survey we consider \cite{Euclid:2019clj,Euclid:2023tqw,Euclid:2023rjj}: a survey area of 15000 deg$^2$;  $N_z=10$ (number of photo-z bins),  $\bar{n}_{gal}=30$ arcmin$^{-2}$ (galaxy number density), $z_m=0.9$, $\gamma=3/2$, $\sigma_\epsilon=0.3$ (intrinsic ellipticity), $\ell_{min}=10$ (minimum multipole) for both GC$_{\rm ph}$ and WL and $\ell_{max}=1500$ (maximum multipole) for WL and  $\ell_{max}=750$ for GC$_{\rm ph}$; for GC$_{\rm sp}$ we adopt $n_z=4$ (number of spectro-z bins), the centers of the bins are $z_i=\{1.0, 1.2, 1.4, 1.65\}$, the error on redshift is 0.001, the minimum scale $k_{\rm min}=0.001$ h\,Mpc$^{-1}$ and the maximum scale is $k_{\rm max}=0.25$ h\,Mpc$^{-1}$. These specifications can be considered very close to the real case. To test the power in constraining of the non-linear scales we also adopt a smaller $k_{\rm max}=0.15$ h\,Mpc$^{-1}$ for GC$_{\rm sp}$ and smaller $\ell_{\rm max}$ respectively $\ell_{\rm max}=500$ for GC$_{\rm ph}$ and $\ell_{\rm max}=1000$ for WL.  We refer to this case as quasi-linear (QL). 
\item for a SKAO-like survey \cite{SKA:2018ckk}: we consider a survey area of 5000 deg$^2$;  $N_z=10$,  $\bar{n}_{gal}=2.7$ arcmin$^{-2}$, $z_m=1.1$, $\gamma=1.25$, $\sigma_\epsilon=0.3$, for $\ell_{min}$ and $\ell_{max}$ as well as for $k_{\rm max}$  we use the same as \textit{Euclid}-like for both spectroscopic and photometric probes. 
\end{itemize}

We use a Fisher matrix approach \cite{Tegmark:1997yq,Seo:2005ys,Seo:2007ns}  to estimate errors for cosmological and model parameter measurements as implemented in the  publicly available library \texttt{CosmicFish} \citep[\href{https://cosmicfish.github.io}{\faicon{github}}]{Raveri:2016xof}. This approach  approximates the curvature of the likelihood at the peak, under the assumption that it is a Gaussian function of the model parameters

For the forecast analysis we will vary the following cosmological parameters around their fiducial values:
\begin{eqnarray}
&&\Omega_{\rm m}=0.2565,\quad \Omega_{\rm b}=0.041,\quad h=0.737\nonumber \\
&& n_{\rm s}=0.96605,\quad \sigma_8=0.891\,,
\end{eqnarray}
where $\Omega_{\rm m}=\Omega_{\rm c}+\Omega_{\rm b}$ and $h=H_0/100$ and the model parameters which fiducial values are:
\begin{equation}
q=1.06, \quad s=0.65.
\end{equation}
The fiducial values have been chosen to be the best fit values for GCCG with  \textit{Planck} data obtained in Ref.~\cite{Frusciante:2019puu}. For stability reason both $\{q,s\}$ need to be positive. However  the use of the Fisher matrix approach to compute the constraints does not allow us to impose \textit{a priori} such cuts in the parameter space. Therefore while the results assume Gaussianity we will \textit{a posteriori} cut the negative range when presenting our results. 

\subsection{Results}

\begin{figure*}[t!]
\includegraphics[width=0.9\textwidth]{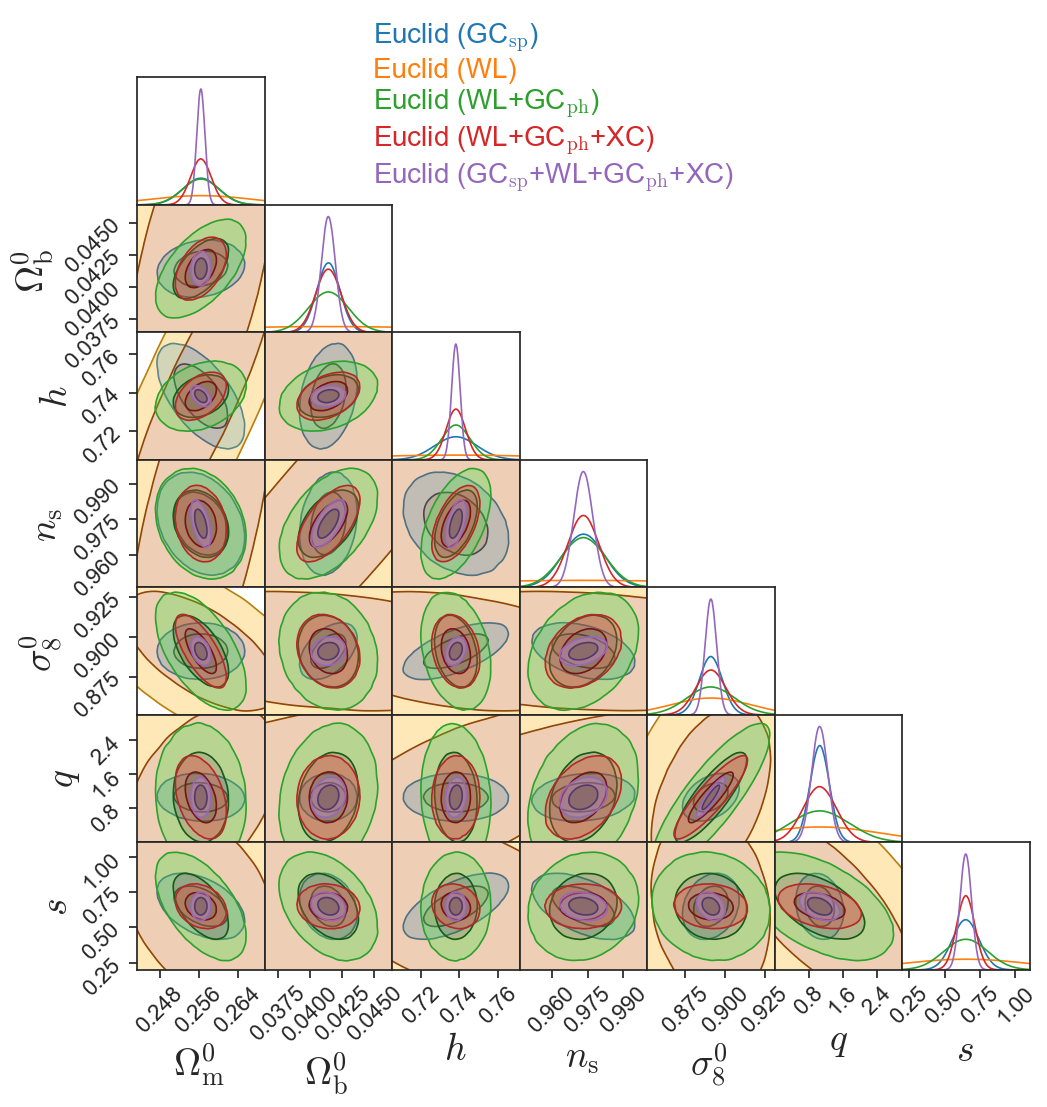}
    \caption{1$\sigma$ and 2$\sigma$ joint marginalized error contours on the cosmological and GCCG model parameters for a \textit{Euclid}-like survey. Blue is GC$_{\rm sp}$, orange is for WL, green  for the combination WL+ GC$_{\rm ph}$,  red  for WL+ GC$_{\rm ph}$+XC and magenta for all the photometric probes, including their cross-correlation, combined with GC$_{\rm sp}$. }
    \label{Fig:Euclid}
\end{figure*}

\begin{figure*}[t!]
\includegraphics[width=0.9\textwidth]{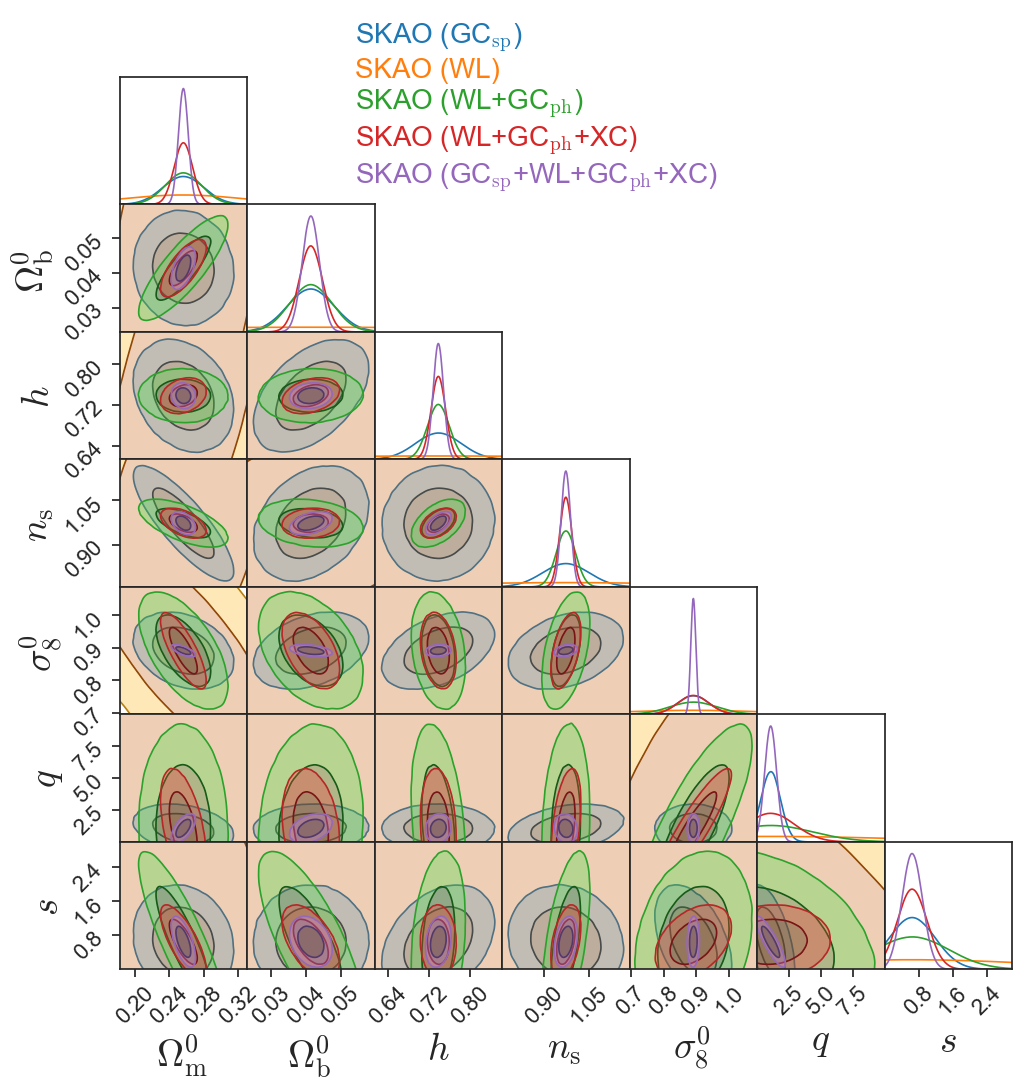}
    \caption{1$\sigma$ and 2$\sigma$ joint marginalized error contours on the cosmological and GCCG model parameters for an SKAO-like survey. Blue is GC$_{\rm sp}$, orange is for WL, green  for the combination WL+ GC$_{\rm ph}$,  red  for WL+ GC$_{\rm ph}$+XC and magenta for all the photometric probes, including their cross-correlation, combined with GC$_{\rm sp}$.}
    \label{Fig:SKAO}
\end{figure*}

\begin{figure*}[t!]
\includegraphics[width=0.9\textwidth]{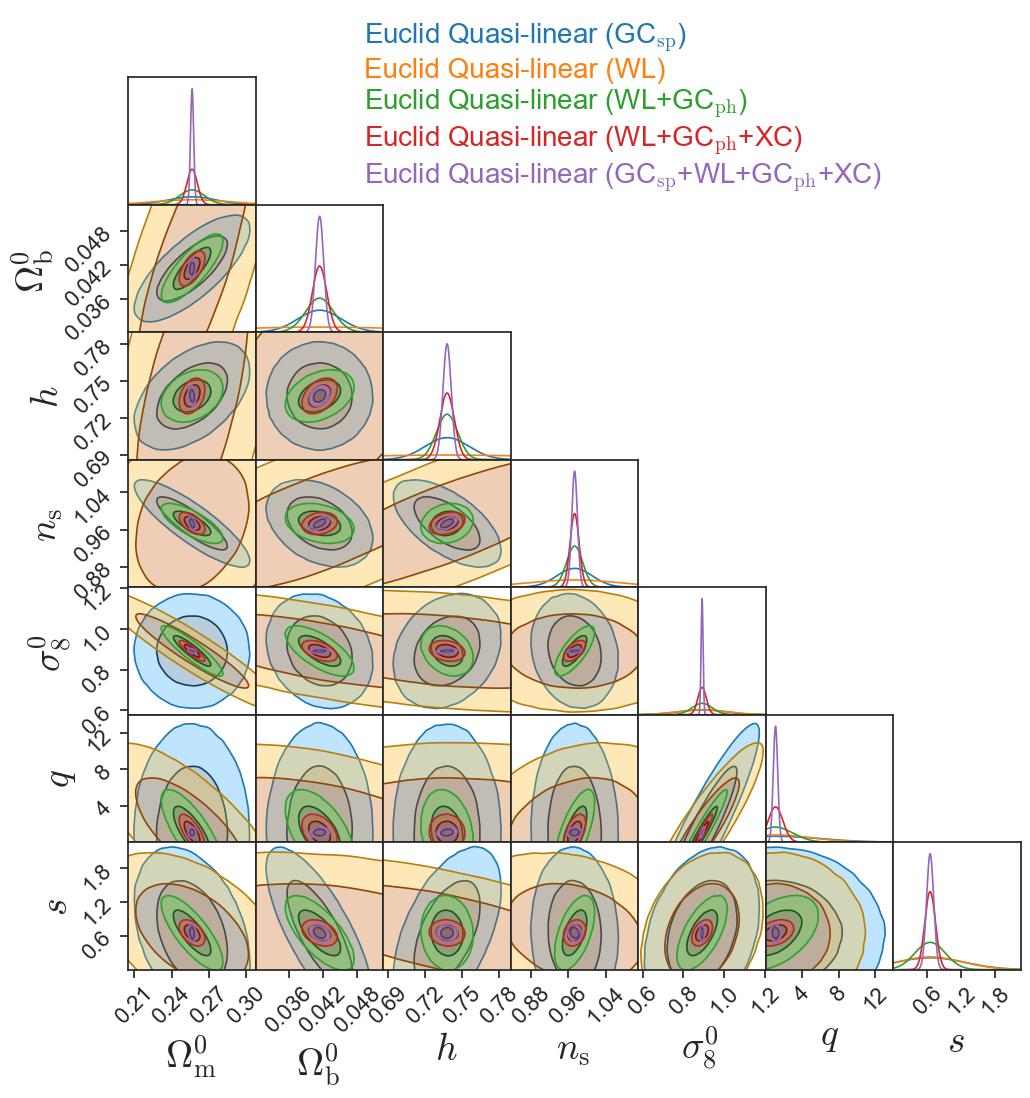}
    \caption{1$\sigma$ and 2$\sigma$ joint marginalized error contours on the cosmological and GCCG model parameters for a \textit{Euclid}-like survey when quasi-linear cuts to linear and angular scales are applied. Blue is GC$_{\rm sp}$, orange is for WL, green  for the combination WL+ GC$_{\rm ph}$,  red  for WL+ GC$_{\rm ph}$+XC and magenta for all combination of probes. }
    \label{Fig:EuclidQL}
\end{figure*}

\begin{figure*}
\includegraphics[width=0.9\textwidth]{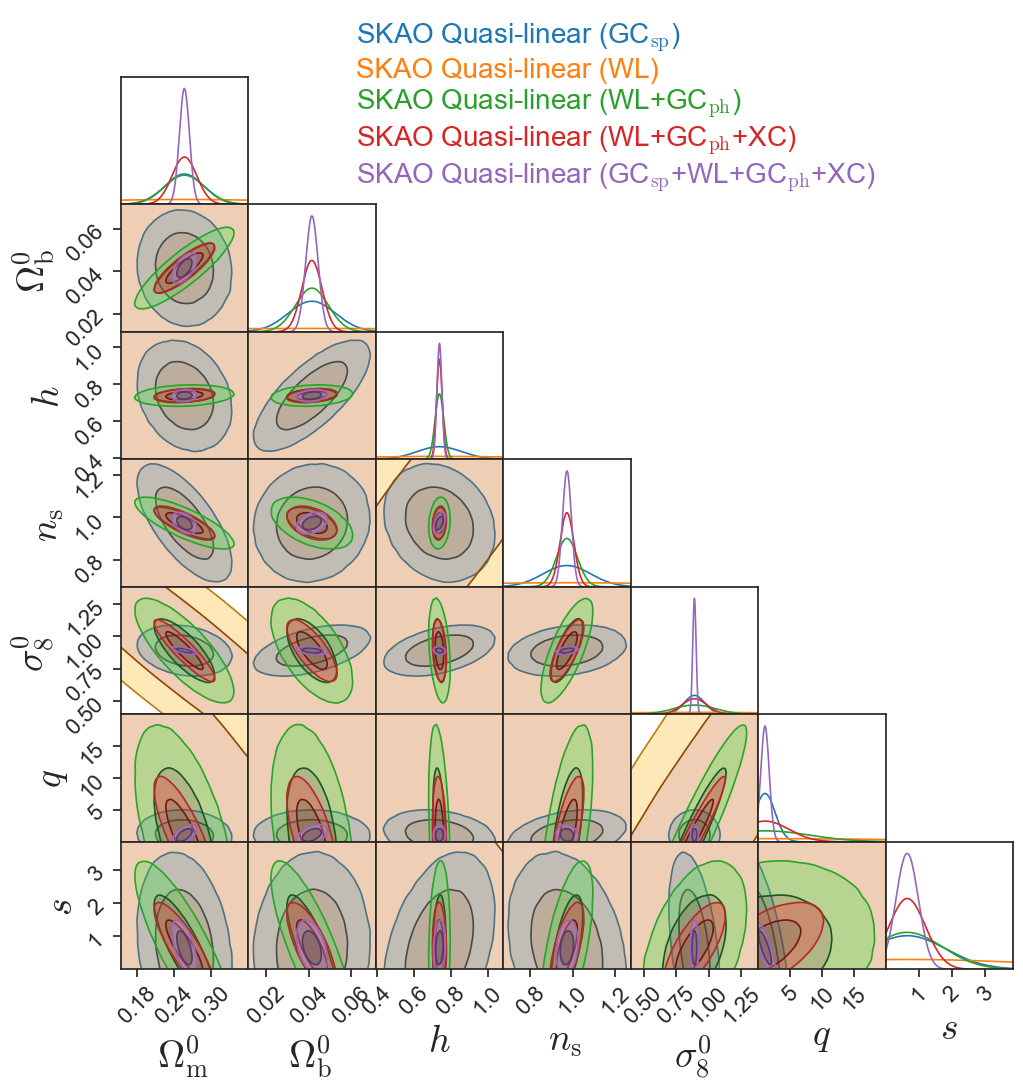}
    \caption{1$\sigma$ and 2$\sigma$ joint marginalized error contours on the cosmological and GCCG model parameters for a SKAO-like survey when quasi-linear cuts to linear and angular scales are applied. Blue is GC$_{\rm sp}$, orange is for WL, green  for the combination WL+ GC$_{\rm ph}$,  red  for WL+ GC$_{\rm ph}$+XC and magenta for all the photometric probes, including their cross-correlation, combined with GC$_{\rm sp}$.}
    \label{Fig:SKAOQL}
\end{figure*}

We show in \autoref{Fig:Euclid} and \autoref{Fig:SKAO} the 1$\sigma$ and 2$\sigma$ contours for the probes GC$_{\rm sp}$, WL,  WL+GC$_{\rm ph}$, WL+GC$_{\rm ph}$+XC and all the combined probes, GC$_{\rm sp}$+WL+GC$_{\rm ph}$+XC, respectively for \textit{Euclid}-like and SKAO-like.  We also show in \autoref{Fig:EuclidQL} and \autoref{Fig:SKAOQL} the  1$\sigma$ and 2$\sigma$ contours for the same combinations of probes in the QL case for both surveys. In  \autoref{Tab:forecasts}  we list the forecasted 1$\sigma$ relative errors to its fiducial on  the cosmological and model's parameters for the same combinations of probes. 

As shown in \autoref{Tab:forecasts}, for the MG parameter $s$  we find that at 1$\sigma$ the \textit{Euclid}-like survey has a stronger power in constraining compared to SKAO-like survey. In details, the GC$_{\rm sp}$ for a \textit{Euclid}-like survey  will constrain $s$ with a relative error of $\sim 14.4\%$, while for SKAO-like survey it is much higher $\sim 82.5\%$; in the WL alone  case the relative error is $\sim 67.1\%$ for \textit{Euclid}-like, while for SKAO-like the parameter $s$ is unconstrained; when the WL is combined with GC$_{\rm ph}$ the relative error decreases to $23.5\%$ for \textit{Euclid}-like, while $s$ is still unconstrained for SKAO-like. The error strongly improves when the XC is considered: $\sim 9.7\%$ and $\sim 53.4\%$ for \textit{Euclid} and SKAO-like respectively. Finally the full combination gives a further better constraint in the case of \textit{Euclid}-like ($\sim 6.2\%$ relative error) and a slightly worse error for SKAO-like, $\sim 36.9\%$. In the SKAO case then the cross-correlation between the photometric probes as well as its combination with spectroscopic GC will be crucial to set a constraint on the $s$ parameter.

When we consider the QL specifications for both surveys the constraints on the $s$ parameter become worse, as it can be expected given that we cut the power in constraining coming from the larger $k$.  Specifically with  the  \textit{Euclid}-like specifications  for the spectroscopic probe we obtain $\sim 93\%$, for weak lensing $\sim 87.4\%$ and for WL combined with GC$_{\rm ph}$ we find $40.6\%$. For SKAO-like these probes alone are not able to constrain $s$. The combinations with the cross correlation of GC$_{\rm ph}$ and WL largely improve the relative error which is comparable to the one with more realistic specifications ($\sim 14.2$ with WL+GC$_{\rm ph}$+XC and $9.6\%$ for the full combination). For SKAO-like we have $84.3\%$ and $51.5\%$.

For the MG parameter $q$,  we also find that at 1$\sigma$ the \textit{Euclid}-like survey performs better compared to the SKAO-like survey. In details, for \textit{Euclid}-like the relative error is $\sim 21.2\%$ with GC$_{\rm sp}$ while for SKAO-like it is $\sim 71.8\%$. For both surveys WL alone is not able to constrain the parameter $q$. In the case of \textit{Euclid}-like the combination of WL+GC$_{\rm ph}$ improves the error which is $65.1\%$, while for SKAO-like the inclusion of the photometric galaxy clustering probe does not make any difference. 
Finally the inclusion of the XC in the photometric probes for the \textit{Euclid}-like survey gives an even  better constraint  $\sim 36.9\%$ and the strongest constraint is $\sim 17.7\%$ for the full combination. For SKAO-like we obtain a constraint $\sim 43.5\%$ only for the full combination. In the QL regime \textit{Euclid}-like loses its power in constraining for both the spectroscopic and photometric GC, while the errors when the XC is included increase ($\sim 75.4\%$ with WL+GC$_{\rm ph}$+XC and $\sim 23\%$ for the full combination), but still performs better than SKAO-like. In the SKAO-like we get a constraint only in the full combination of probes give the orthogonality of the spectroscopic and photometric probes, see  \autoref{Fig:SKAOQL}.

In \autoref{Tab:forecasts} we show also the forecasts on the cosmological parameters $\{\Omega_m^0,\Omega_b^0,h,n_s \sigma_8^0\}$. We can see how the \textit{Euclid}-like survey performs very well on all parameters and for all the combinations of probes considered. The only parameter with a larger relative error is $\Omega_b^0$, $\sim 25.5\%$ for WL alone, but this is usually also the case of $\Lambda$CDM \cite{Euclid:2019clj}. In the SKAO-like case the constraints are all worse with respect to the \textit{Euclid}-like case. In the QL case the constraints degrade for both surveys with \textit{Euclid}-like case performing in any case better than SKAO.
Comparing our forecast on the cosmological parameters with those of the $\Lambda$CDM  obtained for \textit{Euclid} in \cite{Euclid:2019clj} (see Table 9  pessimistic case) we notice that the order of magnitude of the errors is the same as we find in this work, with GCCG performing slightly better on some parameters and for some probes. While this is surprising, because in general the inclusion of additional parameters with respect to the baseline ones leads to larger errors,  we attribute this different tendency here  to the presence of the parameter $s$ at level of the background which affects the constraints on cosmological parameters. Indeed from \autoref{Fig:Euclid} in the bottom line,  we can notice that the orientation of the ellipses of the spectroscopic and photometric probes are such that their combination may help in reducing the errors.

\begin{table*}
\centering
\begin{tabular}{|l||c|c|c|c|c|c|c|}
\hline
 Data &  $\Omega_{\rm m}^0  $ & $\Omega_{\rm b}^0  $ & $h$ & $n_{\rm s}$ & $\sigma_8^0$ & $q $ & $s$\\ \hline \hline
\textit{Euclid} (GC$_{\rm sp}$) & 1.4\% & 2.2\% & 1.5\% & 0.9\% & 0.8\% & 21.2\% & 14.4\%\\
\textit{Euclid} (WL) & 4.0\% & 25.5\% & 7.1\% & 6.7\% & 2.8\% & 132.2\% & 67.1\%\\
\textit{Euclid} (WL+GC$_{\rm ph}$) & 1.4\% & 3.7\% & 1.0\% & 1.0\% & 1.7\% & 65.1\% & 23.5\%\\
\textit{Euclid} (WL+GC$_{\rm ph}$+XC) & 0.8\% & 2.4\% & 0.7\% & 0.7\% & 1.0\% & 36.9\% & 9.7\%\\
\textit{Euclid} (GC$_{\rm sp}$+WL+GC$_{\rm ph}$+XC) & 0.3\% & 1.3\% & 0.3\% & 0.4\% & 0.4\% & 17.7\% & 6.2\%\\
\hline
\textit{Euclid} Quasi-linear (GC$_{\rm sp}$) & 7.2\% & 9.0\% & 2.4\% & 3.8\% & 12.5\% & 449.6\% & 93.0\%\\
\textit{Euclid} Quasi-linear (WL) & 11.6\% & 38.6\% & 11.1\% & 9.9\% & 13.4\% & 370.2\% & 87.4\%\\
\textit{Euclid} Quasi-linear (WL+GC$_{\rm ph}$) & 3.9\% & 5.8\% & 1.1\% & 1.7\% & 5.5\% & 177.0\% & 40.6\%\\
\textit{Euclid} Quasi-linear (WL+GC$_{\rm ph}$+XC) & 1.6\% & 3.0\% & 0.8\% & 1.0\% & 2.3\% & 75.4\% & 14.2\%\\
\textit{Euclid} Quasi-linear (GC$_{\rm sp}$+WL+GC$_{\rm ph}$+XC) & 0.5\% & 1.7\% & 0.4\% & 0.6\% & 0.5\% & 23.0\% & 9.6\%\\
\hline
SKAO (GC$_{\rm sp}$) & 9.1\% & 16.0\% & 6.0\% & 7.9\% & 5.3\% & 71.8\% & 82.5\%\\
SKAO (WL) & 27.2\% & 146.2\% & 46.0\% & 42.8\% & 24.3\% & 920.8\% & 440.5\%\\
SKAO (WL+GC$_{\rm ph}$) & 8.1\% & 14.6\% & 2.9\% & 3.3\% & 8.1\% & 307.2\% & 131.6\%\\
SKAO (WL+GC$_{\rm ph}$+XC) & 4.1\% & 8.0\% & 1.9\% & 2.1\% & 5.3\% & 176.5\% & 53.4\%\\
SKAO (GC$_{\rm sp}$+WL+GC$_{\rm ph}$+XC) & 2.2\% & 5.9\% & 1.4\% & 1.6\% & 0.9\% & 43.5\% & 36.9\%\\
\hline
SKAO Quasi-linear (GC$_{\rm sp}$) & 11.9\% & 26.4\% & 16.4\% & 11.5\% & 8.8\% & 147.2\% & 177.6\%\\
SKAO Quasi-linear (WL) & 74.8\% & 239.8\% & 69.7\% & 59.2\% & 84.4\% & 2246.8\% & 605.0\%\\
SKAO Quasi-linear (WL+GC$_{\rm ph}$) & 12.4\% & 18.5\% & 3.2\% & 5.0\% & 18.0\% & 646.3\% & 161.3\%\\
SKAO Quasi-linear (WL+GC$_{\rm ph}$+XC) & 7.6\% & 11.3\% & 2.1\% & 3.3\% & 10.7\% & 343.7\% & 84.3\%\\
SKAO Quasi-linear (GC$_{\rm sp}$+WL+GC$_{\rm ph}$+XC) & 3.1\% & 7.0\% & 1.8\% & 2.1\% & 1.4\% & 61.7\% & 51.5\%\\
\hline
\end{tabular}
\caption{Marginalized 1$\sigma$ relative errors on the cosmological and GCCG model parameters using \textit{Euclid}-like and SKAO-like specifications for spectroscopic galaxy clustering (GC$_{\rm sp}$), weak lensing (WL), photometric galaxy clustering (GC$_{\rm ph}$) and the cross-correlation between the photometric probes (XC).}
\label{Tab:forecasts}
\end{table*}

\section{Conclusion} \label{Sec:Conclusion}

In this work we have presented the non-linear modeling of the matter power spectrum in the Generalized Cubic Covariant Galileon with the aim of exploiting the constraining power of future surveys.  To this extent, we created a new patch for the \texttt{ReACT} code which implements the halo-model reaction prescription. For our model  we implemented the modified background evolution, the spherical collapse and  the potential energy obtained from the virial theorem. Additionally we included the 1-loop corrections to compute the 1-loop matter power spectra.  We demonstrated the performance of this new tool by comparing our theoretical prediction against $N$-body simulations and we found  agreement within 5\%. 

Finally we focused on the ability of future missions such as \textit{Euclid} and SKAO to constrain the GCCG model parameters. 
We found that a survey such as \textit{Euclid} will be able to  provide outstanding constraints on the two parameters of the model $\{q,s\}$, especially when a  larger range of non-linear scales is considered.  We also provide constraints for SKAO which are weaker for both parameters when compared to \textit{Euclid}.

In conclusion, with the tool presented in this work   it will be possible to obtain reliable and accurate constraints on the GCCG model with forthcoming surveys. We are confident that when the new data will be available it will allow to discern the GCCG model from $\Lambda$CDM. 

\acknowledgments
We thank M. Martinelli and F. Pace  for useful discussion.
L.A. is supported by Fundação para a Ciência e a Tecnologia (FCT) through the research grants UIDB/04434/2020, UIDP/04434/2020 and  from the FCT PhD fellowship grant with ref. number 2022.11152.BD.
N.F. is supported by the Italian Ministry of University and Research (MUR) through the Rita Levi Montalcini project ``Tests of gravity on cosmic scales" with reference PGR19ILFGP.
L.A. and N.F.  also acknowledge the FCT project with ref. number PTDC/FIS-AST/0054/2021 and  the COST Action CosmoVerse, CA21136, supported by COST (European Cooperation in Science and Technology).
BB is supported by a UKRI Stephen Hawking Fellowship (EP/W005654/2).
For the purpose of open access, the author(s) has applied a Creative Commons Attribution (CC BY) licence to any Author Accepted Manuscript version arising.
\bibliography{bib}

\end{document}